\documentclass[12pt]{article}

\usepackage{
graphicx,amsmath,amssymb,bm}

\newtheorem{prop}{Prop}[section]
\newtheorem{lemma}[prop]{Lemma}

\newtheorem{theorem}[prop]{Theorem}

\title{
General properties of overlap operators \\ in disordered
quantum spin systems}
\author{C. Itoi \\ 
Department of Physics, GS $\&$ CST, Nihon University, \\
Kanda-Surugadai, Chiyoda, Tokyo 101-8308, Japan} 
\begin{document}
\maketitle
\begin{abstract}
We study short-range  quantum spin systems with Gaussian disorder.
We obtain quantum mechanical extensions of the Ghirlanda-Guerra identities. 
We discuss properties of overlap spin operators with these identities.\\

 \noindent
 {\bf keywords} quenched disorder $\cdot$  spin glass $\cdot$ quantum spin systems $\cdot$ Ghirlanda-Guerra identities
 \end{abstract}
 \section{Introduction}
\label{intro}
In this paper, we
study short-range quantum spin models 
with random interactions in $d$-dimensional cubic lattice 
$\Lambda_L= {\mathbb Z}^d \cap [1,L]^d$. To define short-range interactions, we construct a collection $C_L$ of interaction ranges in the following way. 
Let $m$ be an arbitrarily fixed positive integer independent of $L$ and $X_k$ $(k=1,2, \cdots, m)$ be  a subset of $\Lambda_L$ which contains  $(1,1, \cdots, 1) \in \Lambda_L$, such that
the distance $|i-j|$ between any two sites $i,j \in X_k$  has an upper bound independent of $L$ for each $k$. 
We define  $C_L$ by the collection of all translated subsets of $X_1, \cdots, X_m$
\begin{equation}
C_L:=\bigcup_{k=1}^m \{X_k +i  \subset \Lambda_L |  i \in  \Lambda_L\}.
\end{equation}
For example, as $C_L$ we can employ  $\Lambda_L$ itself or nearest neighbor bonds 
$$B_L = \{ \{i, j\}|  i,j \in \Lambda _L, |i-j|=1\}.$$
The spin operator $\sigma_i^\mu \ (\mu=x,y,z)$ at a site $i \in \Lambda_L$ acting on the Hilbert space $\bigotimes_{j \in \Lambda_L} {\cal H}_j$
is defined by a tensor product of the Pauli matrix $\sigma^\mu$ acting on ${\cal H}_i \simeq {\mathbb C}^2$  and two dimensional identity matrices. 
For an arbitrary  $X \in C_L$, we denote 
$$
\sigma_X ^\mu = \prod_{i \in X} \sigma_i^\mu.
$$  
We define a Hamiltonian  as a function of spin operators and
i.i.d. standard Gaussian random variables  $(g_X)_{X \in C_L}$
\begin{equation}
H_{L}(\sigma,g):=-J \sum_{X \in C_L} \sum_{\mu=x,y,z} g_X ^{\mu} \sigma_X^\mu + H_{\rm non}(\sigma),
\label{hamil}
\end{equation}
where  $J$ is a real constant and $H_{\rm non}(\sigma)$ is an arbitrary  non-random short-range Hamiltonian.
We assume that $H_{\rm non}(\sigma)$ is defined also by another collection $C'_L$ of interaction ranges
as in the same way to define the random Hamiltonian.  
\\

\noindent
{\it Examples}  \\
1. Random  field Heisenberg model \\
For $C_L=\Lambda_L$ and the Heisenberg model  Hamiltonian $ H_{\rm non} (\sigma),$
the Hamiltonian becomes
\begin{equation} 
H_{L}(\sigma,g)=-J \sum_{i \in \Lambda_L} g_i  \sigma_i^\mu -\sum_{X \in B_L} \sum_{\nu = x,y,z} \sigma_X^\nu.
\end{equation}
2. Random bond Heisenberg model 
\\
For $C_L=B_L$ and $H_{\rm non} (\sigma)=\sum_{X \in B_L} \sum_{\mu=x,y,z}J g_0 \sigma_X^\mu$ with a real constant $g_0$, the Hamiltonian becomes
\begin{equation}
H_{L}(\sigma,g)=-J \sum_{X \in B_L} \sum_{\mu=x,y,z} (g_X ^{\mu} + g_0 )\sigma_X^\mu.
\end{equation}
This bond randomness corresponds to a non-centered Gaussian disorder.
\\
3. Other models\\
The Hamiltonian (\ref{hamil}) contains some other physically interesting models, such as Heisenberg model with random next nearer neighbor
interactions,  and 
with random plaquette  interactions. \\

For a positive $\beta $ and a real number $J$,  the  partition function is defined by
\begin{equation}
Z_L(\beta, J,g) := {\rm Tr} e^{ - \beta H_L(\sigma,g)},
\end{equation}
where the trace is taken over the Hilbert space.
The expectation of an arbitrary function $f$ 
of spin operators with respect to the Gibbs measure is denoted by
\begin{equation}
\langle f(\sigma) \rangle := \frac{1}{Z_L(\beta,J,g)}{\rm Tr} f(\sigma)  e^{ - \beta H_L(\sigma,g)}.
\end{equation}

We define the following functions of  $(\beta, J) \in [0,\infty) \times {\mathbb R}$ and $(g_X^\mu)_{X \in C_L, \mu=x,y,z}$
\begin{equation}
\psi_L(\beta, J,g) := \frac{1}{|\Lambda_L|\beta} \log Z_L(\beta,J,g), \\ 
\end{equation}
$-\psi_L(\beta,J,g)|\Lambda_L |$ is called free energy in statistical physics.
We define a function $p_L:[0,\infty) \times {\mathbb R} \rightarrow {\mathbb R}$ by
\begin{eqnarray}
p_L(\beta,J):={\mathbb E} \psi_L(\beta,J,g) ,
\end{eqnarray}
where ${\mathbb E}$ stands for the expectation of the random variables $(g_X^\mu)_{X \in C_L, \mu=x,y,z}$.
Here, we introduce a fictitious time  $t \in [0,\beta]$ and define a time evolution of operators with the Hamiltonian.
Let $O$ be an arbitrary self adjoint operator, and we define an operator valued function  $\hat O$ of $t$ by
\begin{equation}
\hat  O(t):= e^{-tH} O  e^{tH}.
\end{equation}
Furthermore, we define the  Duhamel expectation of time depending operators 
$\hat O_1(t_1),  \cdots, \hat O_k(t_k)$  by
$$
(\hat O_1 \hat O_2\cdots  \hat O_k)_{\rm D} := \int_{[0,\beta]^k} dt_1 dt_2\cdots dt_k \langle {\rm T} [\hat O_1(t_1) \hat O_2(t_2) \cdots \hat O_k(t_k)] \rangle,
$$
where the symbol ${\rm T}$ is a multilinear mapping of the chronological ordering.
If we define a partition function with arbitrary self adjoint operators  $O_0, O_1, \cdots, O_k$ and real
numbers $x_1, \cdots, x_k$
$$
Z(x_1,\cdots, x_k) := {\rm Tr} \exp \beta \left[O_0+\sum_{i=1} ^k x_i O_i \right],
$$
the Duhamel expectation of $k$ operators represents
 the $k$-th order derivative of the partition function 
 \cite{Cr,S}
$$(\hat O_1\cdots \hat O_k)_{\rm D}=\frac{1}{Z}
\frac{\partial ^k Z}{\partial x_1 \cdots \partial  x_k}.
$$
Here, we consider a term of random Hamiltonian per interaction range
as a function of a sequence $\sigma^\mu=(\sigma^\mu_X)_{ X \in C_L }$ of spin operators and 
the random variables $(g_X^\mu)_{X \in C_L}$ with an arbitrarily fixed index $\mu$
by 
$$
h_L(\sigma^\mu,g^\mu)
 :=\frac{1}{|C_L|} \sum_{X \in C_L} g_X ^\mu \sigma_X^\mu.
 $$
 We denote its time evolution by
 $$
\hat h_L ^\mu(t) := h_L(\hat \sigma^\mu(t),g^\mu)
.$$
 The covariance of these operators with the expectation in $g$ 
 \begin{equation} {\mathbb E}( h_L (\sigma_a^\mu,g^\mu) h_L (\sigma_b^\mu,g^\mu)
 )
 =  |C_L|^{-1} R(\sigma_a^\mu,
 \sigma_b^\mu )\end{equation}
where the overlap $R(\sigma^\mu _a, \sigma_b^\mu)$  is defined by
$$
R(\sigma_a^\mu,\sigma_b^\mu):=\frac{1}{|C_L|} \sum_{X \in C_L}\sigma^\mu_{X,a}\sigma^\mu_{X,b}.
$$
For example, in the random field Heisenberg model, 
this becomes the site overlap operator
$$R(\sigma_a^\mu,\sigma_b^\mu)=\frac{1}{|\Lambda_L|} \sum_{i \in \Lambda_L}\sigma^\mu_{i,a}\sigma^\mu_{i,b}.
$$
In the random bond Heisenberg model, it becomes the bond overlap operator
$$
R(\sigma_a^\mu,\sigma_b^\mu)=\frac{1}{|B_L|} \sum_{X \in B_L}\sigma^\mu_{X,a}\sigma^\mu_{X,b}= 
\frac{1}{|B_L|} \sum_{\{i,j\} \in B_L} \sigma^\mu_{i,a}  \sigma^\mu_{j,a}  \sigma^\mu_{i,b}   \sigma^\mu_{j,b}.
$$
In the short-range spin glass models, the bond overlap is independent of the site overlap
unlike the Sherrington-Kirkpatrick (SK) model \cite{SK}, where the bond-overlap is identical to the square of the site-overlap. 
We denote its time evolution with $(t_a,t_b) \in [0,\beta]^2$
$$
\hat R_{a,b}^\mu:= R_{a,b}(\hat \sigma_a^\mu(t_a),\hat \sigma_b^\mu(t_b))
$$
Let $f$ be a polynomial  of $x_1, \cdots, x_n$
$$
f(x_1, \cdots, x_n) = \sum_{i_1, \cdots, i_n} C(i_1, \cdots, i_n ) x_1^{i_1} \cdots x_n^{i_n},
$$
and we define  its Duhamel expectation of $n$-replicated time depending spin operators  by
preserving its multilinear property   
$$
(\hat f)_{\rm D}:=(f(\hat \sigma_{i_1,a_1}^{\mu_1}, \cdots, \hat \sigma_{i_n,a_n}^{\mu_n} ) )_{\rm D} 
=\sum_{i_1, \cdots, i_n} C(i_1, \cdots, i_n )
(\hat \sigma_{i_1,a_1}^{\mu_1} \cdots \hat \sigma_{i_n,a_n}^{\mu_n} )_{\rm D}.
$$
The overlap is a polynomial of two replicated time depending spin operators. 
The Duhamel expectation of the  overlap for two different replicas  is identical to its  normal expectation by the Gibbs measure
$$
(\hat R_{1,2}^\mu )_{\rm D} =\frac{1}{|C_L|} \sum_{X \in C_L}(\hat \sigma_{X,1}^\mu \hat \sigma_{X,2}^\mu 
)_{\rm D}
=\frac{1}{|C_L|} \sum_{X \in C_L}\langle  \sigma_{X,1}^\mu \sigma_{X,2}^\mu 
 \rangle = 
  \langle R^\mu_{1, 2}\rangle.
  $$
  Hereafter, we consider $\hat h^\mu_L$ and $\hat R_{a,b} ^\mu$ with an  arbitrarily fixed index $\mu$,  then
  $\hat h_L$ and 
  $\hat R_{a,b}$ denote them without the index $\mu$ for simpler notation. 
  
In this paper, we prove two main theorems with respect to the overlap operators.
 
{\theorem  \label{c1}  Let $n$ be a positive integer and let $f$ be a bounded polynomial function of $n$ 
replicated time depending spin operators $\hat \sigma_{i,a}^{\mu} (t_a) $.  For any $(\beta,  J) \in [0,\infty) \times {\mathbb R}$, we have
\begin{eqnarray}
\hspace{-5mm}\lim_{L \rightarrow \infty } {\mathbb E}\left[\sum_{a=1} ^n(  (
 \hat R_{1,a}- 
  \hat R_{1,n+1}-
   \hat R_{a,n+1} )\hat f)_{\rm D}
 +
 ((n+1) \hat R_{1,2} -\hat R_{1,1})_{\rm D} ( \hat f )_{\rm D}\right]=0. \label{GG1} 
 \end{eqnarray}}
 
{\theorem \label{MT}  Let $n$ be a positive integer and let $f$ be a bounded polynomial function of $n$ 
replicated time depending spin operators $\hat \sigma_{i,a}^{\mu} (t_a) $. There is a mesure zero set
$A \subset [0,\infty) \times {\mathbb R}$,  such that  
for any $(\beta, J) \notin A$ we have 
 \begin{eqnarray}
\lim_{L \rightarrow \infty }\left[\sum_{a=1} ^n {\mathbb E}  (
 (\hat R_{1,a} 
- \hat R_{1,n+1})\hat f)_{\rm D} +{\mathbb E} (\hat R_{1,2}  -\hat R_{1,1})_{\rm D}{\mathbb E} (\hat  f )_{\rm D}   \right]=0. \label{GG2}
 \end{eqnarray}
 }\\
Theorem \ref{c1} gives  quantum mechanical extensions of identities obtained by Contucci and Giardin\`a \cite{CG,CG2,CG3},
and  Theorem \ref{MT}  gives those of the Ghirlanda-Guerra identities \cite{GG}.
In classical disordered Ising spin systems, the overlap $R^z_{a,b}$  satisfies the Ghirlanda-Guerra identities which are valid universally \cite{GG,T}. 
Sometimes, these identities are called Aizenman-Contucci identities \cite{AC,CG3}. 
The Ghirlanda-Guerra identities are useful to understand properties of
the random variables $R_{a,b}$.  In classical  mean field spin glass models such as 
the Sherrington-Kirkpatrick model \cite{SK} and mixed $p$-spin models, 
it is well-known that the distribution of the overlap shows
broadening for sufficiently large $\beta$. 
This phenomenon is called replica symmetry breaking,
conjectured by Parisi \cite{Pr} and proved by Talagrand \cite{T2,T} for the SK model. 
The replica symmetry breaking is observed  generally in mean field spin glass models \cite{Pn}.
Originally, the Ghirlanda-Guerra identities were obtained to understand the ultrametricity of replicated spin configurations in mean field spin glass models. This has been proved by Panchenko using the Ghirlanda-Guerra identities \cite{Pa2}.
On the other hand, recently, Chatterjee has proved that the random variable $R_{a,b}$  is single valued in the random field Ising model in any dimensions at any temperature \cite{C2}. 
This result implies that the replica symmetry breaking does not occur in that model.
The proof  is constructed by utilising  the Fortuin-Kasteleyn-Ginibre inequality \cite{FKG} and the Ghirlanda-Guerra identities. 
Quantum mechanical  extensions of these identities are expected to be useful to study the properties of  overlap operators in quantum spin glass models as well. We study general properties of the expectation of overlap operators with these identities. 
There are several  approaches to obtain the Ghirlanda-Guerra identities \cite{AC,C2,C1,CG3,GG,Pn,T}. 
 Here, we employ a similar approach to Chatterjee's method \cite{C2,C1}.


\section{Proof}
{\lemma \label{lim3} For every $(\beta,J) \in [0,\infty) \times {\mathbb R}$,  the following limit exists
$$
p(\beta,J) = \lim_{L \rightarrow \infty} p_L(\beta, J).
$$
Proof.}  This is proved by a standard argument based on the decomposition of the lattice  into disjoint blocks \cite{CGP,CL,Gff}. 
 Fix $(\beta, J) \in [0,\infty) \times {\mathbb R}$, and let $L, M$ be positive integers and denote  $N=LM$, then we divide the lattice $\Lambda_N$ into $M^d$ disjoint translated blocks of 
 $\Lambda_L$.
 Define a new Hamiltonian $H$ on $\Lambda_N$ by deleting interaction ranges in $C_{\rm del}$
 near the boundaries of blocks from $C_L$ and $C'_L$,
 such that $M^d$ spin systems on the blocks have no  interaction with each other. 
 The original Hamiltonian $H_N$ has the following two terms
 $$
 H_N=H+H_{\rm del},
 $$
 where $H_{\rm del}$ consists of the terms of interaction ranges in $C_{\rm del}$. 
 We consider the following function of $x$ defined by
 $$
 \varphi_N(x) :=\frac{1}{|\Lambda_N|\beta} {\mathbb E}\log {\rm Tr} \exp[-\beta (H+ x H_{\rm del})].
 $$
 Not that $\varphi_N(1)=p_N(\beta, J)$ and $\varphi_N(0)= p_L(\beta, J).$ 
 The derivative functions of $\varphi_N$ are given by
 $$
 \varphi_N'(x) =-\frac{1}{|\Lambda_N|}  {\mathbb E} \langle H_{\rm del} \rangle_x, \ \ \ 
 \varphi_N''(x) =\frac{\beta }{|\Lambda_N|}  {\mathbb E} (\delta \hat H_{\rm del} \delta \hat H_{\rm del} )_{{\rm D}, x} \geq 0,
  $$
 where 
 $\langle A\rangle_x$ and
  $(AB)_{{\rm D},x}$ are  the Gibbs and the Duhamel expectations  respectively 
  with the Hamiltonian $H+xH_{\rm del}$ for operators  $A$ and $B$, and 
  $\delta\hat H_{\rm del} =  \hat H_{\rm del} -\langle H_{\rm del} \rangle_x$.  
  The second inequality gives
 $$
 \varphi_N(1) \geq \varphi_N(0) +\varphi_N'(0) = \varphi_N(0) - \frac{1}{|\Lambda_N|}  {\mathbb E}  \langle H_{\rm del}\rangle_0.
 $$
 In the same argument for $\varphi_N(1-x)$, we obtain  
 $$
 \varphi_N(0) \geq \varphi_N(1) -\varphi_N'(1) = \varphi_N(1) + \frac{1}{|\Lambda_N|}  {\mathbb E}  \langle H_{\rm del} \rangle_1
 $$
 Therefore 
 $$
 p_L(\beta, J) - \frac{1}{N^d}  {\mathbb E} \langle H_{\rm del} \rangle_0 
\leq  p_N(\beta, J) \leq  p_L(\beta, J) - \frac{1}{N^d}  {\mathbb E}  \langle H_{\rm del} \rangle_1
 $$
 Since the spin operator is bounded, the expectation of $H_{\rm del}$ is bounded by the number of interactions $|C_{\rm del}|$. Therefore,
there exist  positive numbers $K_1$ and $K$ independent of $L$ and $N$, such that the function $p_N$ and $p_L$ obey  
$$
|p_N-p_L| \leq\frac{K_1 |C_{\rm del}|}{N^d} \leq \frac{Kd}{L}.
$$
In the same argument for $M$ instead of $L$, we have
$$
|p_{N}-p_M|  \leq \frac{K'd}{M},
$$
and therefore 
$$
|p_L-p_{M}| \leq |p_N-p_L| +|p_{M}-p_N| \leq \frac{Kd}{L}+\frac{K'd}{M},
$$
The sequence $p_L$ is Cauchy. $\Box$

 \paragraph{Note} Functions $p,$ $\psi$, $p_L$ and $\psi_L$ are convex functions of each argument for other arbitrarily fixed arguments, since every 
 second order derivative in each argument is the corresponding Duhamel expectation between the same operators, which is nonnegative.  \\

We define a set $A$ by a set of $(\beta,J) \in [0,\infty) \times {\mathbb R}$ where  $p$ is not differentiable. 

\paragraph{Note} The set $A$ is measure zero, since the function $p$ is convex in each argument for every fixed another argument.  
\\

Hereafter, we concern about dependence of $\psi_L$ only on $(g_X^\mu)_{X \in C_L}$ with the  fixed $\mu$, 
then we  use a lighter notation $\psi_L(g^\mu)=\psi_L(\beta,J,g)$.  
We  define a generating  function $\gamma_L(u)$ of a parameter $u \in [0,1] $ by
\begin{equation}
\gamma_L(u) :=|\Lambda_L| {\mathbb E} ({\mathbb E}_1\psi_L( \sqrt{u} g^\mu+ \sqrt{1-u}g^\mu_1) )^2,
\end{equation}
where 
${\mathbb E }$ denotes the expectation in all random variables,
and  ${\mathbb E }_1$ denotes the  expectation
in only $g_1^\mu$. This generating function $\gamma_L$ is  introduced by Chatterjee \cite{C}.

{\lemma \label{varp} 
For any $(\beta, J) \in [0,\infty) \times {\mathbb R}$, any positive integer  $L$ and  any $u \in [0,1]$, each order derivative of the function $\gamma_L$ is represented by  
\begin{eqnarray}
\gamma_L^{(n)}(u) &=& |\Lambda_L| \beta^{n} J^n \sum_{X_1 \in C_L}  \cdots \sum_{X_n \in C_L}
{\mathbb E} \left({\mathbb E}_1\psi_{L,X_n,\cdots,X_1}(\sqrt{u} g^\mu+ \sqrt{1-u}g_1^\mu ) \right)^2 ,
\end{eqnarray}
where we denote 
$$\psi_{L,X_n,\cdots, X_1} (g^\mu) := \frac{\partial^n \psi_L (g^\mu )}{\partial g_{X_1}^\mu \cdots \partial g_{X_n}^\mu }.
$$
There exists a positive constant $K$, such that 
for any positive integer $n$ and for any $u_0 \in [0,1)$, an upper bound on the $n$-th order derivative of the function $\gamma_L$
is given by 
\begin {equation}
\gamma_L^{(n)}(u_0) \leq \frac{(n-1)! K J^2}{(1-u_0)^{n-1}}.
\end{equation}

Proof. }A straightforward calculation of derivatives and integration by parts in the 
Gaussian integral yield the identity.  The non-negativity of the derivatives in an arbitrary order   
guarantees that $\gamma^{(n)}_L(u) $ is monotonically increasing in $u$. 
Also note that $\gamma_L^{(n)}(u)$ is bounded, as far as the system size $L$ is finite. 
The first derivative of $\gamma_L$ is bounded by 
$$
\gamma_L'(1)  = |\Lambda_L| \sum_{X \in C_L} {\mathbb E} \left(\psi_{L,X}( g^\mu) \right)^2
=\frac{J^2}{|\Lambda_L|} \sum_{X \in C_L} {\mathbb E} \langle \sigma_X ^\mu \rangle^2 \leq K J^2,
$$ 
where $K > 0 $ is a constant independent of $L$.
We have used $E \langle \sigma_X ^\mu \rangle^2 \leq 1$.  
The function $\gamma_L(u)$ is continuously differentiable any times in $t$ for finite $L$.
From Taylor's theorem, for any positive $\beta$, $J$,  any integer $n \geq 1$ and any $u_0 \in [0,1)$, there exists $u_1 \in (u_0,1)$ such that
$$
\gamma'_L(1) = \sum_{k=0} ^{n-1} \frac{(1-u_0)^k}{k!} \gamma_L^{(k+1)}(u_0)+ \frac{(1-u_0)^{n}}{n!}  \gamma_L^{(n+1)}(u_1).
$$ 
For any $u_0 \in [0,1)$, each term in right hand side is bounded by  $\gamma'_L(1)$,  and therefore
$$
\frac{(1-u_0)^{n-1}}{(n-1)!} \gamma_L^{(n)}(u_0) \leq  \gamma'_L(1) \leq  KJ^2.
$$
 This completes the proof. $\Box$
\\

\paragraph{Note.} To evaluate the variance of the function $\psi_L(\beta,J,g)$, 
we define another generating function for all random variables $(g^\mu_X)_{X,C_L,\mu=x,y,z}$
\begin{equation}
\chi_L(u) :=|\Lambda_L| {\mathbb E} ({\mathbb E}_1\psi_L(\beta, J,  \sqrt{u} g+ \sqrt{1-u}g_1) )^2.
\end{equation}
The same  result as Lemma \ref{varp} also for $\chi_L(u)$ gives
the following upper bound  on the variance of the function $\psi_L$ is obtained
\begin{equation}
|\Lambda_L|Var(\psi_L)=\chi_L(1)-\chi_L(0)=\int_0^1du \chi_L'(u)  \leq \chi_L'(1)   \leq 3K J^2.
\label{varineq}
\end{equation}
The similar result is obtained in \cite{Cr}.

 Here,  we define two types of deviations of an arbitrary operator $\hat O$ by
$$
\delta \hat O:=\hat O -(\hat O )_{\rm D}, \
 \ \  \Delta \hat O := \hat O- 
 {\mathbb E} 
 (\hat O ) _{\rm D} .
 $$
We prove the following two lemmas for these deviations of $\hat h_L$.

{\lemma  \label{delta} 
For any $(\beta,J) \in [0,\infty) \times {\mathbb R}$, we have
\begin{equation}
\lim_{L\rightarrow \infty} {\mathbb E}(\delta \hat h_L\delta \hat  h_L )_{\rm D}=0.
\label{a}
\end{equation}
Proof. }
We calculate the following Duhamel expectation 
of $\delta \hat h_L= \hat h_L -\langle h_L \rangle$ with integration by parts, then we use the Cauchy-Schwarz
inequality and Lemma {\ref{varp}}
\begin{eqnarray}
&&{\mathbb E} (\delta \hat h_L  \delta \hat h_L) _{\rm D}=\frac{1}{|C_L|^2} {\mathbb E} \sum_{X,Y \in C_L} g_X^\mu g_Y^\mu (\delta \hat \sigma_X^\mu  \delta \hat \sigma_Y^\mu )_{\rm D}
=\frac{1}{|C_L|^2}\sum_{X,Y \in C_L}  {\mathbb E} \left(\frac{\partial^2}{\partial g_X \partial g_Y}+ \delta_{X,Y}
 \right)(\delta \hat \sigma_X ^\mu \delta \hat \sigma_Y^\mu)_{\rm D} \nonumber  \\
&& = \frac{|\Lambda_L|}{|C_L|^2\beta J^2}\left(  \sum_{X,Y ,Z,W \in \Lambda_L} \delta_{X,Z} \delta_{Y,W} {\mathbb E} \psi_{L,W,Z,Y,X}(g)
+ \sum_{X,Y \in C_L}  \delta_{X,Y} {\mathbb E} \psi_{L,Y,X}(g)
 \right)\nonumber\\
&&\leq  \frac{|\Lambda_L|}{|C_L|^2\beta J^2}\Big( \sqrt{\sum_{X,Y ,Z,W \in C_L} \left({\mathbb E} \psi_{L,W,Z,Y,X}(g)
\right)^2\sum_{X,Y,Z,W\in C_L} (\delta_{X,Z} \delta_{Y,W} )^2}  \nonumber  \\
&&\hspace{3cm} +\sqrt{\sum_{X,Y \in C_L} \left({\mathbb E}\psi_{L,Y,X}(g)
\right)^2\sum_{X,Y \in C_L} \delta_{X,Y} ^2}  \ \Big) \nonumber \\
&&= \frac{|\Lambda_L|}{ |C_L|^2 \beta J^2}\left(
\sqrt{\frac{|C_L|^2 \gamma_L^{(4)}(0)}{|\Lambda_L|\beta^4J^4} }+
\sqrt{\frac{|C_L|  \gamma_L^{(2)}(0)}{|\Lambda_L| \beta^2J^2}} \ \right) 
\leq  \frac{\sqrt{K}}{\beta^2 J^2}\left(\sqrt{\frac{6|\Lambda_L|}{\beta J|C_L|^2}} +\sqrt{ \frac{|\Lambda_L|}{|C_L|^3}}\right).
\end{eqnarray}
Since we have $C |\Lambda_L| \leq |C_L | \leq C'| \Lambda_L|$ for some $C,C'$ independent of $L$,  this
 gives the limit (\ref{a}).
$\Box$
\\

{\lemma \label{lem} For any $(\beta, J) \notin A$, we have
\begin{equation}
\lim _{L \rightarrow \infty} {\mathbb E} ( \Delta \hat h_L \Delta \hat h_L
 )_{\rm D}=0.
 \label{lim2}
\end{equation}

\noindent
Proof.} 
Note the relations
$$ \langle h_L\rangle = \frac{\partial \psi_L}{\partial J}, \hspace{1cm}
{\mathbb E} \langle h_L\rangle = \frac{\partial p_L}{\partial J}. $$
For an arbitrary $\epsilon >0 $, the convexity of $\psi_L$ implies
$$ \frac{\psi_L(J)-\psi_L(J-\epsilon)}{\epsilon}\leq\frac{\partial \psi_L }{\partial J}(J) \leq \frac{\psi_L(J+\epsilon )-\psi_L(J)}{\epsilon}.$$
Here we use  an indicator projection  $I$ defined by $I[true]=1$ and $I[false]=0$ .  For $(\beta, J) \notin A$ and $\epsilon > 0$, 
the convexity of $\psi_L$ implies 
\begin{eqnarray}
&&{\mathbb E} \Big( \langle h_L\rangle-\frac{\partial p}{\partial J}\Big)^2  \nonumber \\
&&={\mathbb E} \Big( \left(\frac{\partial \psi_L }{\partial J}  -\frac{\partial p}{\partial J}\right)^2 \left(I\left[ \frac{\partial \psi_L }{\partial J}  \geq \frac{\partial p}{\partial J}\right] +I\left[ \frac{\partial \psi_L }{\partial J}  < \frac{\partial p}{\partial J}\right] \right)\Big) \nonumber  \\
&&\leq {\mathbb E}\left( \Big(\frac{\psi_L(J+\epsilon)-\psi_L(h)}{\epsilon}  -\frac{\partial p}{\partial J}\Big)^2 I\left[ \frac{\partial \psi_L }{\partial J}  \geq \frac{\partial p}{\partial J}\right]  \right)
\nonumber \\ 
&&\hspace{1cm}+ {\mathbb E} \left(\Big(\frac{\psi_L(J)-\psi_L(J-\epsilon)}{\epsilon} - \frac{\partial p}{\partial J}\Big)^2
I \left[ \frac{\partial \psi_L }{\partial J}  < \frac{\partial p}{\partial J}\right] \right)\nonumber  \\
&&\leq{\mathbb E} \Big( \frac{\psi_L(J+\epsilon)-\psi_L(J)}{\epsilon}  -\frac{\partial p}{\partial J}\Big)^2+
{\mathbb E} \Big( \frac{\psi_L(J)-\psi_L(J-\epsilon)}{\epsilon}  -\frac{\partial p}{\partial J}\Big)^2,
\end{eqnarray}
Furthermore, the inequality (\ref{varineq}) gives
$
{\mathbb E} (\psi_L-p_L)^2  \leq \frac{3 J^2}
{|\Lambda_L|},
$
which yields the following bound
\begin{eqnarray}
&&{\mathbb E} \Big(  \frac{\psi_L(J+\epsilon)-\psi_L(J)}{\epsilon}-\frac{\partial p}{\partial J}(J) \Big)^2 \nonumber \\
&&\leq \frac{3(2|J|+\epsilon)^2}{\epsilon^2|\Lambda_L|} +\Big(\frac{p_L(J+\epsilon) -p(J+\epsilon)}{\epsilon}   
- \frac{p_L(J) -p(J)}{\epsilon}   + \frac{p(J+\epsilon)-p(J)}{\epsilon}-\frac{\partial p}{\partial J}(J)  \Big)^2 .\nonumber
\end{eqnarray}
In the limit $L \rightarrow \infty$, Lemma \ref{lim3} implies 
$$
\lim_{L\rightarrow \infty}{\mathbb E} \Big( \langle h_L\rangle -\frac{\partial p}{\partial J}(J) \Big)^2
\leq\Big( \frac{p(J+\epsilon)-p(J)}{\epsilon}-\frac{\partial p}{\partial J}(J)  \Big)^2 
+ \Big( \frac{p(J)-p(J-\epsilon)}{\epsilon}-\frac{\partial p}{\partial J}(J)  \Big)^2$$
Since this bound holds for any  $\epsilon >0 $
and $p(J)$ is differentiable at $J$,  we have 
$$
\lim_{L\rightarrow \infty}{\mathbb E} 
\Big( \langle h_L\rangle -\frac{\partial p}{\partial J}(J) \Big)^2 =0
$$
The above  and the identity  (\ref{a}) yield the identity (\ref{lim2}), 
since
\begin{eqnarray}
{\mathbb E} (\Delta \hat h_L \Delta \hat h_L  )_{\rm D}  = 
{\mathbb E}(\delta \hat h_L\delta \hat h_L)_{\rm D}+{\mathbb E}\Big(\langle  h_L \rangle-\frac{\partial p}{\partial J}(J) \Big)^2
- \Big(\frac{\partial p}{\partial J}(J) -{\mathbb E} \langle h_L\rangle \Big)^2. \nonumber  
\end{eqnarray}
$\Box$ \\

Here, we prove the main theorems.
\paragraph{ Proof of Theorem \ref{c1}.}
In the following expectation of the energy density,
the  integration by parts gives
\begin{equation}
{\mathbb E} ( \hat h_L \hat f )_{\rm D} =\frac{1}{|C_L|} \sum_{X \in C_L} {\mathbb E} g_X^\mu (\hat \sigma^\mu_X\hat f )_{\rm D}
= \frac{1}{|C_L|} \sum_{X \in C_L} {\mathbb E} \frac{\partial}{\partial g_X^\mu} ( \hat \sigma_X  ^\mu\hat f )_{\rm D}\\
=\beta J \sum_{a=1}^n {\mathbb E} ( (\hat R_{1,a}-\hat R_{1,n+1})\hat  f )_{\rm D} ,
\label{correq}
\end{equation}
and  also we have
\begin{eqnarray}
{\mathbb E}\langle h_L\rangle(\hat f)_{\rm D}&=&\frac{1}{|C_L|} \sum_{X \in C_L} {\mathbb E} g_X^\mu  \langle \sigma_X^\mu\rangle (\hat f)_{\rm D}=\frac{1}{|C_L|} \sum_{X \in C_L} {\mathbb E} \frac{\partial}{\partial g_X^\mu} \langle \sigma_X^\mu\rangle (\hat f)_{\rm D}
 \nonumber \\
&=&\frac{\beta J}{|C_L|} \sum_{X \in C_L} {\mathbb E}[ ((\hat \sigma_X^\mu\hat \sigma_X ^\mu)_{\rm D} - \langle \sigma_X^\mu\rangle^2)(\hat f)_{\rm D}+ 
\langle \sigma_X ^\mu \rangle\sum_{a=1}^n( (\hat \sigma^\mu_{X,a}\hat f)_{\rm D}- \langle \sigma_X^\mu\rangle (\hat f)_{\rm D}) ] \nonumber  \\
&=& \beta J {\mathbb E}[((R_{1,1})_{\rm D} - (n+1)\langle R_{1,2}\rangle )(\hat f)_{\rm D}+\sum_{a=1}^n(\hat R_{a,n+1}\hat f)_{\rm D}].\end{eqnarray}
Therefore 
$$
{\mathbb E} (\delta  \hat h_L \hat f )_{\rm D} =\beta J \left[\sum_{a=1} ^n( {\mathbb E}  (
\hat R_{1,a} \hat f )_{\rm D}-{\mathbb E} (  \hat R_{1,n+1} \hat f)_{\rm D} -{\mathbb E} (  \hat R_{a,n+1} \hat f)_{\rm D})
-{\mathbb E} [( \hat R_{1,1}-(n+1) \hat R_{1,2})_{\rm D} ( \hat f )_{\rm D}]\right]
$$
The absolute value of the left hand side is bounded by
$$
|{\mathbb E}( \delta \hat h_L  \hat f )_{\rm D}| \leq \sqrt{{\mathbb E}(\delta \hat h_L\delta \hat h_L)_{\rm D}{\mathbb E}(\hat f\hat f)_{\rm D}}.
$$
This and Lemma \ref{delta} give the identity (\ref{GG1}). $\Box$

\paragraph{ Proof of Theorem \ref{MT}.} 
Under the condition of Lemma \ref{lem}, we show the  identity (\ref{GG2}). First we substitute $f=1$ into the identity (\ref{correq}), then we have
\begin{equation}
{\mathbb E}\langle h_L\rangle=\frac{\beta J}{|C_L|} \sum_{X \in C_L}
{\mathbb E}[ (\hat \sigma_X^\mu \hat \sigma_X^\mu )_{\rm D}- \langle\sigma_X^\mu \rangle^2 ]=\beta J {\mathbb E} (\hat R_{1,1}-\hat R_{1,2})_{\rm D}.
\label{hL}\end{equation}
Therefore, 
$$
{\mathbb E}( \Delta \hat h_L\hat f )_{\rm D}=\beta J\left[\sum_{a=1} ^n {\mathbb E}(  \hat R_{1,a}\hat f)_{\rm D} -
n {\mathbb E} (  \hat R_{1,n+1} \hat f )_{\rm D} - {\mathbb E}(\hat R_{1,1}-\hat R_{1,2})_{\rm D} {\mathbb E} (\hat f)_{\rm D} \right].
$$
The absolute value of the left hand side is bounded by
$$
|{\mathbb E}( \Delta \hat h_L \hat  f )_{\rm D} |\leq \sqrt{{\mathbb E}(\Delta \hat h_L\Delta \hat h_L)_{\rm D}{\mathbb E}(\hat f \hat f)_{\rm D}}
$$
By the limit (\ref{lim2}) in Lemma \ref{lem}, this  converges to zero,
 then this completes the proof  of Theorem \ref{MT}.
$\Box$
\\
\section{Discussions}
Here we discuss general properties of overlap operators with the obtained identities.
The identity (\ref{GG1}) in Theorem \ref{c1}  for $n=1$ and $\hat f=\hat R_{1,1}$ gives 
$$
\lim_{L \rightarrow \infty }[ {\mathbb E} (\hat R_{1,1}\hat R_{1,1})_{\rm D} -{\mathbb E} (\hat R_{1,1})_{\rm D} ^2
- 2{\mathbb E} (\hat R_{1,2} \hat R_{1,1})_{\rm D}+2 {\mathbb E}  \langle 
 R_{1,2}\rangle( \hat R_{1,1})_{\rm D}  ]=0.
$$
The identities (\ref{GG2})  in Theorem \ref{MT} for $n=1$,  $\hat f=\hat R_{1,1}$
, for $n=2,$ $\hat f=\hat R_{1,2}$ and for $n=3,$ $\hat f=\hat R_{2,3}$ are
\begin{eqnarray}
&&\lim_{L \rightarrow \infty }[ {\mathbb E} (\hat R_{1,1}\hat R_{1,1})_{\rm D} -({\mathbb E} (\hat R_{1,1})_{\rm D} )^2
- {\mathbb E} ( \hat R_{1,2} \hat R_{1,1})_{\rm D} +
{\mathbb E}  \langle 
  R_{1,2}\rangle {\mathbb E} ( \hat R_{1,1})_{\rm D}  ]=0,\\
  &&\lim_{L \rightarrow \infty }[ {\mathbb E} (\hat R_{1,1}\hat R_{1,2})_{\rm D} 
  +{\mathbb E} (\hat R_{1,2}\hat R_{1,2})_{\rm D}-2{\mathbb E} (\hat R_{1,3}\hat R_{1,2})_{\rm D}
  -({\mathbb E} (\hat R_{1,1})_{\rm D} - {\mathbb E}  \langle 
  R_{1,2}\rangle) {\mathbb E} \langle  R_{1,2}\rangle ]=0, \nonumber \\
&&\lim_{L \rightarrow \infty }[ {\mathbb E} (\hat R_{1,1}\hat R_{2,3})_{\rm D} +2 {\mathbb E} ( \hat R_{1,2} \hat R_{1,3})_{\rm D}-3{\mathbb E} \langle R_{1,2}\rangle ^2
 -({\mathbb E} ( \hat R_{1,1})_{\rm D} -
{\mathbb E}  \langle 
  R_{1,2}\rangle ){\mathbb E}  \langle 
  R_{1,2}\rangle ]=0 \nonumber , 
\end{eqnarray}
The above four identities give us
\begin{eqnarray}
\hspace{-1cm}&&
{\mathbb E} ( \delta \hat R_{1,2} \delta \hat R_{1,1})_{\rm D} 
=
\frac{1}{2}{\mathbb E} ( \delta \hat R_{1,1} \delta  \hat R_{1,1})_{\rm D},  \\
\hspace{-1cm}&&
{\mathbb E} ( \Delta \hat R_{1,2} \Delta \hat R_{1,1})_{\rm D}
 =
 {\mathbb E} ( \Delta \hat R_{1,1} \Delta  \hat R_{1,1})_{\rm D}, \\
\hspace{-1cm}&&
{\mathbb E} ( \delta \hat R_{1,3} \delta \hat R_{1,2})_{\rm D} 
=
\frac{1}{4}{\mathbb E} ( \delta \hat R_{1,2} \delta  \hat R_{1,2})_{\rm D} +
 \frac{1}{8}{\mathbb E} ( \delta \hat R_{1,1}   \delta \hat R_{1,1})_{\rm D} \\
\hspace{-1cm}&&
3{\mathbb E}  \langle 
 R_{1,2}\rangle^2-  {\mathbb E} ( \hat R_{1,2}  \hat R_{1,2})_{\rm D}-2 ( {\mathbb E} \langle R_{1,2} \rangle )^2   
 =
 \frac{3}{2}{\mathbb E} (  \delta  \hat R_{1,1}  \delta \hat R_{1,1})_{\rm D}
+2{\mathbb E}( \Delta  \hat R_{1,1})_{\rm D} ^2 ,
\end{eqnarray}
in the limit $L \rightarrow \infty$.
Note that all are nonnegative.
The third identity gives 
\begin{equation}
\lim_{L \rightarrow \infty}
{\mathbb E} ( \delta \hat R_{1,3} \delta \hat R_{1,2})_{\rm D} \geq \lim_{L \rightarrow \infty}
\frac{1}{4}{\mathbb E} ( \delta \hat R_{1,2} \delta  \hat R_{1,2})_{\rm D}.
\end{equation}
The final one gives some inequalities
\begin{equation}
0 \leq \lim_{L \rightarrow \infty} \frac{3}{2}
 {\mathbb E}(\delta \hat R_{1,2} \delta \hat R_{1,2})_{\rm D}
\leq \lim_{L \rightarrow \infty} {\mathbb E}( \Delta \hat R_{1,2}  \Delta \hat R_{1,2})_{\rm D} \leq
\lim_{L \rightarrow \infty}3 {\mathbb E}\langle \Delta   R_{1,2} \rangle^2.
\end{equation}
These inequalities become equalities in the classical limit, since $\hat R_{1,1}=1$ because of the commutativity between spin operators and the Hamiltonian.


\end{document}